\documentclass[aps,twocolumn,superscriptaddress,showpacs,prl]{revtex4}
\usepackage{bm}
\usepackage{amssymb}
\usepackage{graphicx,epsfig}
\usepackage{subfigure}

\begin{document}
\title{Macroscopic Many-Qubit Interactions in Superconducting Flux Qubits}
\author{Sam Young \surname{Cho}}
\email{sycho@physics.uq.edu.au}
\affiliation{
 Centre for Modern Physics and Department of Physics, Chongqing University, Chongqing 400044, China}
\affiliation{
Department of Physics, The University of Queensland, 4072, Australia}
\author{Mun Dae \surname{Kim}}
\email{mdkim@kias.re.kr}
\affiliation{
Korea Institute for Advanced Study, Seoul 130-722, Korea}
\date{\today}

\begin{abstract}
 Superconducting flux qubits
 are considered to investigate macroscopic many-qubit interactions.
 Many-qubit states based on current states can be manipulated
 through the current-phase relation in each superconducting loop.
 For flux qubit systems comprised of $N$ qubit loops,
 a general expression of low energy Hamiltonian is presented
 in terms of low energy levels of qubits and macroscopic quantum
 tunnelings between the many-qubit states.
 Many-qubit interactions classified by {\em Ising type-
 or tunnel-}exchange interactions can be observable
 experimentally.
 Flux qubit systems can provide various artificial-spin systems
 to study  many-body systems that cannot be found naturally.
\end{abstract}

\maketitle

{\it Introduction.}$-$
 The last decade has seen rapidly developing advanced material technologies
 that make it possible to investigate previously inaccessible quantum systems
 for quantum information and computation in solid-state systems.
 Especially, coherent manipulation of quantum states
 in tunable superconducting devices has enabled to demonstrate
 macroscopic qubits
 \cite{ChargeQ,Mooij99,Yu}
 and entangled states of qubits
 \cite{Pashkin,Izmalkov,Berkley}.
 Experimentally, it has been shown that,
 in terms of pseudo-spins,
 different types of exchange interactions between two artificial-spins  such as
 an Ising interaction
 for charge qubits \cite{Pashkin}  and flux qubits \cite{Izmalkov,Majer}
 and
 an XY interaction
 for phase qubits \cite{Berkley},
 can be realized and controlled by the system parameters.

 It is believed that electrons are interacting in a pair.
 The interaction is called two-body interaction.
 Normally, in a number of spin systems such as spin chains and lattices,
 the two-body interactions in the spin pairs reveal rich
 many-electron physics.
 Understanding the many-electron effects
 is one of the most important researches in condensed matter physics.
 Additionally, in a strongly correlated electronic system,
 a low energy spin Hamiltonian can involve  more than three
 spin interactions \cite{Thouless,Roger,MacDonald}.
 Such  multiple spin interactions are known to play a significant role
 for quantum phase transitions.
 However,
 multiple artificial-spin interactions are not yet investigated,
 although artificial-spin exchange (two-body) interactions are demonstrated
 in different types of solid-state qubit systems.
 This work aims to discuss, in a general framework,
 how artificial-multiple spin interactions are possible and
 realizable in superconducting qubit systems.
 In fact, flux qubit systems are shown to have an intrinsic property
 which is multiple artificial-spin interactions.
 Accordingly, flux qubit systems enable to study
 various artificial-spin systems corresponding to
 many-body systems unlikely found naturally.
%

 In a superconductor, the macroscopic wavefunction
 can be written by
 $\psi({\bf r})=\sqrt{n^*} \, e^{i \varphi({\bf r})},$
 where $n^*$ and $\varphi({\bf r})$ are
 the density and phase of Cooper pairs, respectively.
 $\psi({\bf r})$ describes the behavior of
 the entire ensemble of Cooper pairs in the superconductor.
 The supercurrent density in electromagnetic field is given by
\begin{equation}
{\bf J} = \frac{q^* n^*}{m^*}  \Big(\hbar\nabla \varphi({\bf r})
             -q^* {\bf A}({\bf r}) \Big),
 \label{Js}
\end{equation}
 where $q^*$ and $m^*$ are respectively the charge and mass of Cooper pairs.
 Then, the current states of flux qubit loops are influenced by the
 variations of the phase $\varphi({\bf r})$ across Josephson junctions
 and the vector potential ${\bf A}({\bf r})$.
 A change of current state in a qubit loop results in a change
 of current states in other qubit loops
 because (i) the change of Josephson junction phases
 in superconducting loops coupling qubits
 mediates the change of the currents states of all qubits and
 (ii) the circulating current in the qubit produces
 the induced magnetic flux that influences on all other qubits.
 In experiments, several ways to make two- or four-flux qubits interacting
 have been employed.
 Disconnected supconducting loops, as the indirect way, are coupled inductively
 \cite{Izmalkov,Majer}
 by means of the induced magnetic flux.
 Other direct ways are to introduce
 connecting superconducting loops \cite{Kim,Kim2006,Ploeg},
 which is called phase coupling.
 Consequently, many flux qubits defined by current states
 can interact all together, which can be observable in experiments.

 We present a general expression of $N$-qubit Hamiltonian
 describing low energy physics.
 The Hamiltonian is determined by the low level
 energies and the tunneling amplitudes between $N$-qubit states
 in the flux qubit systems.
 We define two types of many-qubit exchange interactions originating
 from the energy differences of many-qubit states and
 the macroscopic quantum tunneling between the states.
 Further, it is shown that
 a specific coupling scheme enables to map flux qubit systems
 into many-body systems.

 {\em Model}.$-$
 We consider a general model
 including the inductive and phase coupling ways.
 The $N$ flux qubit systems are composed of
 $N$ qubit loops with $N'$ loops connecting the qubit loops.
 Primed (unprimed) indices will indicate qubit (connecting) loops.
 The charging energy of Josephson junctions in
 the $N(N')$ qubit (connecting) loops is given by
 \begin{equation}
 {\cal H}_C=\! \frac{1}{2}\! \left(\frac{\Phi_0}{2\pi}\right)^2
     \!\!\bigg(
          \sum^N_{i=1}\!\sum_{\alpha} 
           C^\alpha_{i}\dot\varphi^2_{i\alpha}
     \!   +\! \sum^{N'}_{i'=1}\!\!\sum_{\alpha'}
          C^{\alpha'}_{i'}\dot\varphi^2_{i'\alpha'} \bigg),
 \label{CE}
 \end{equation}
 where $C(C')$ are the capacitance of the Josephson junctions
 in the qubit (connecting) loops.
 $\Phi_0 = h/2e$ is the unit flux quantum.
 $\alpha(\alpha')$ rely on the number of Josephson junctions
 in the qubit (connecting) loops.
 The inductive energy is given by
 \begin{eqnarray}
  {\cal H}_L  &=&
  \sum_{i,j=1}^N
  \frac{1}{2}\left(L^{(ij)}+\delta_{ij} L^{(i)}_K \right) I_i I_j
       + \sum_{i,i'=1}^{N,N'} {\cal L}^{(ii')} I_i I'_{i'}
  \nonumber \\
       &&
  +\sum_{i',j'=1}^{N'}
   \frac{1}{2}\left(L'^{(i'j')}+\delta_{i'j'} L'^{(i')}_K \right)
   I'_{i'} I'_{j'},
 \label{LE}
 \end{eqnarray}
 where $I_i(I'_{i'})$ are the circulating currents
 in the qubit (connecting) loop $i(i')$.
 $L^{(ii)}=L^{(i)}_S$ is
 the self-inductance for the qubit loop $i$.
 For $i\neq j$, $L^{(ij)}$ is
 the mutual inductance between the qubits $i$ and $j$.
 $L^{(i)}_K$ is the kinetic-inductance \cite{Kim,Bloch,MajerAPL} in the qubit loop $i$.
 Similarly, $L'^{(i')}_S$, $L'^{(i')}_K$, and $L'^{(i'j')}$
 are denoted for the connecting loops.
 ${\cal L}^{(ii')}$ is
 the mutual inductance between the qubit loop $i$ and the connecting loop $i'$.
 Finally, the Josephson energy of the junctions is given by
 \begin{equation}
 {\cal H}_J\!=\! \sum^{N}_{i=1}\! \sum_{\alpha} 
          2 E^\alpha_{Ji} \sin^2\frac{\varphi^\alpha_i}{2}
    +\sum^{N'}_{i'=1}\sum_{\alpha'}
           2 E'^{\alpha'}_{Ji'} \sin^2\frac{\varphi^{\alpha'}_{i'}}{2},
 \label{JE}
 \end{equation}
 where $E_J$'s are the Josephson energy of junctions
 in the qubit and connecting loops.

 {\it Fluxoid quantization.}$-$
 By integrating Eq. (\ref{Js}) along the closed path in the $i$-th loop,
 the fluxoid quantization gives the boundary conditions,
 \begin{equation}
  L^{(i)}_K I_i/\Phi_0 =
  n_i - \frac{1}{2\pi}\sum^{}_{\alpha} \varphi^{\alpha}_i - f_i ,
 \label{fluxoid}
 \end{equation}
 where $\varphi^\alpha_i$ is the phase
 across the Josephson junction $\alpha$,
 $n_i$ is an integer, and
 $f_i=f^{(i)}_\mathrm{ext} + f^{(i)}_\mathrm{ind}$
 consists of an external and induced magnetic fields, i.e.,
 $f^{(i)}_{\mathrm{ext}}=\Phi_i/\Phi_0$ and
 $f^{(i)}_\mathrm{ind}=\sum^{N}_{j=1} L^{(ij)} I_j/\Phi_0
          +\sum^{N'}_{i'=1} {\cal L}^{(ii')} I'_{i'}/\Phi_0$.
 Similarly, the boundary conditions in the connecting loops can be given.
 From the boundary conditions,
 the total energy can be reexpressed as a function of the phases, $\{\varphi_i\}$
 and their time derivatives, $\{\dot\varphi_i\}$.

 {\em $N$-qubit Hamiltonian}.$-$
 The number of Cooper pairs, $n$, and
 the phase of wavefunction, $\varphi$,
 are non-commuting variables, i.e., $[\varphi,n]=i$,
 such that the canonical momentum, $P_\varphi$, can be introduced
 as  $P_\varphi = n \hbar = -i\hbar\partial_\varphi$ \cite{Mooij94},
 where $n=q/2e$ with the charge from the Josephson relation,
 $q=C (\Phi_0/2\pi)\dot\varphi$.
 When the charging energy is much smaller than the Josephson energy,
 the phase is well-defined while the number is strongly fluctuating.
 The charging energy ${\cal H}_C(\{\dot \varphi_i\})$
 plays a role of kinetic energy for a particle
 in an effective potential
 defined by $U(\{\varphi_i\})={\cal H}_L(\{\varphi_i\})+{\cal H}_J(\{\varphi_i\})$.

 In the three-Josephson junction qubit loops ($\alpha\in\{a,b,c\}$)
 with $E^{b,c}_{Ji}=E_{Ji}$ and $\varphi^{b,c}_{i}=\varphi_i$,
 the effective potential has the $2^N$ local minima corresponding to
 the $2^N$ basis, $\{ \left|m_1, \cdots, m_N\right\rangle \}$ ,
 of the $N$ qubits
 with $m_i = \uparrow$ and $\downarrow$ for $i=1, \cdots, N$.
 The values of
 $\{\varphi_{i}\}$ at the local minimum corresponding to the state
 $\left|m_1,\cdots, m_N\right\rangle$ are denoted by
 $\{\varphi^0_{i;m_1\cdots m_N}\}$.
 Then, $\{\varphi^0_{i;m_1\cdots m_N}\}$ determines
 the current state of flux qubit $i$ by the current-phase relation.

 In the low energy limit,
 one can employ a tight-binding approximation
 in which the $2^N$ states of $N$ qubits correspond to $2^N$-lattice sites.
 In the $2^N$ basis,
 $\{ \left|m_1, \cdots, m_N\right\rangle \}$,
 the low energy $N$ qubit-Hamiltonian matrix
 can be written as
 \begin{equation}
  {\cal H}_{N}
   =     
      \sum_{j_1,\cdots, j_N \in \{0,x,y,z\}}
       C_{j_1\cdots j_N}
      \; \mbox{\boldmath $\sigma$}^{j_1}_1 \otimes
         \cdots \otimes
         \mbox{\boldmath $\sigma$}^{j_N}_N,
 \label{Nqubit}
 \end{equation}
 where
 $\mbox{\boldmath $\sigma$}^{0}(\mbox{\boldmath $\sigma$}^{x,y,z})$
 are the identity (Pauli) matrices.
 The coefficients are obtained by
 \begin{equation}
       C_{j_1\cdots j_N}
 =\frac{1}{2^N} \mathrm{Tr}\left[
       \mbox{\boldmath $\sigma$}^{j_1}_1
       \otimes \cdots \otimes
       \mbox{\boldmath $\sigma$}^{j_N}_N \; {\cal H}_N
        \right].
 \end{equation}
 The diagonal components of the Hamiltonian matrix
 are the level energies, $E_{m_1,\cdots,m_N}$,
 at the local minima, $\{\varphi^0_{i;m_1\cdots m_N}\}$.
 The level energies are given by
 \begin{equation}
 E_{m_1,\cdots, m_N}
  = \frac{\hbar}{2} \sum_{i=1}^N \omega_{i;m_1,\cdots,m_N}
    + U(\{\varphi^0_{i;m_1\cdots m_N}\}),
 \end{equation}
 where
 the characteristic oscillating frequencies
 are
 $\omega^2_{i;m_1,\cdots,m_N}=
 \frac{1}{M_i} \frac{\partial^2}{ \partial\varphi^2_{i}}
 U(\{\varphi_i\}) |_{\{\varphi^0_{i;m_1,\cdots,m_N}\}}$
 with an effective mass
 $M_i = \left(\frac{\Phi_0}{2\pi}\right)^2 C^{(i)}_{\rm eff}$
 and  effective capacitance $C^{(i)}_{\rm eff}$
 in the harmonic oscillator approximation \cite{Orlando}.

 Generally,
 the macroscopic tunneling processes between any two many-qubit
 states are possible due to the quantum fluctuation originating
 from the kinetic energy.
 The off-diagonal components are the macroscopic quantum tunneling amplitudes, i.e.,
 \begin{equation}
   t\, \, : \,
     \left| m'_1, \cdots, m'_N\right\rangle
     \Longleftrightarrow
     \left|m_1, \cdots, m_N\right\rangle
 \end{equation}
 for the tunneling between the two states,
     $\left| m'_1, \cdots, m'_N\right\rangle$ and
     $\left|m_1, \cdots, m_N\right\rangle$.
 The tunneling amplitudes
 can be calculated by the well-known numerical methods such as
 WKB approximation, instanton method, 
 and Fourier grid Hamiltonian method \cite{Kim03}. 
 The tunneling process,
 $\left|\uparrow \uparrow \uparrow \cdots \uparrow \right\rangle
  \Longleftrightarrow
 \left|\downarrow \uparrow \uparrow \cdots \uparrow \right\rangle$,
 describes the first pseudo-spin flip.
 Such a tunneling process that describes
 only one pseudo-spin flip among the $N$ qubits
 is called {\em single qubit tunneling}, $t_{1}$.
 If the $N$ qubits are flipped for tunneling, 
 the tunneling processes can be called {\em N-qubit tunneling}, $t_{N}$,
 e.g.,
 $\left|\downarrow \uparrow\uparrow \cdots \downarrow \right\rangle
  \Longleftrightarrow
  \left|\uparrow \downarrow\downarrow \cdots \uparrow \right\rangle$.
 Normally, single qubit tunneling amplitudes are much larger
 than other multiple qubit ones.
 However,
 when a multiple qubit tunneling amplitude is larger than
 single qubit one, the multiple qubit tunneling processes can play an
 important role in determining the property of eigenstates of the system
 \cite{Kim06,Kim07}.

 {\em Many-qubit interaction.}$-$
 Actually,
 Eq. (\ref{Nqubit}) describes
 any $N$ qubit system including all types of many-qubit interactions.
 Let us expand the low energy $N$ qubit-Hamiltonian matrix
 in terms of qubit interactions;
 \begin{equation}
  {\cal H}_{N}\!
   =\! H_0\!
    + \! \sum_{i} H^{(i)}_1
    +\sum_{i<j} H^{(ij)}_2
    +\!\! \sum_{i<j<k} H^{(ijk)}_3
    +\cdots
    + H^{(1\cdots N)}_N,
 \end{equation}
 where $H_0 = (1/2^N)\mathrm{Tr}\left[ {\cal H}_N\right]$
 and
 the qubits are described by
      $H^{(i)}_1 =
          \varepsilon_{i} \ \mbox{\boldmath $\sigma$}^{z}_i
       + t^{(i)}_1 \mbox{\boldmath $\sigma$}^{x}_i$
 with the energy difference $2\varepsilon_i$
 and the tunneling amplitude $t^{(i)}_1$ between the two states
 of the qubit $i$.
 Qubit interactions are denoted by
 two-qubit interactions
 $H^{(ij)}_2$, three-qubit interactions $H^{(ijk)}_3$, and so on.
 Then,
 the $N$-qubit interaction is presented by
 \begin{equation}
 H^{(1\cdots N)}_{N}
   =      
      \sum_{j_1,\cdots,j_N\in \{x,y,z\}}
       C_{j_1\cdots j_N}
      \; \mbox{\boldmath $\sigma$}^{j_1}_1 \otimes
         \cdots \otimes
         \mbox{\boldmath $\sigma$}^{j_N}_N .
 \end{equation}
 We define the {\em $N$-qubit exchange coupling constant}
 as
 \begin{equation}
  J^{(N)}_{z\cdots z}
  = C_{z\cdots z}
  = \frac{1}{2^N} \mathrm{Tr}\left[
       \mbox{\boldmath $\sigma$}^{z}_1 \otimes
       \cdots \otimes
       \mbox{\boldmath $\sigma$}^{z}_N \; {\cal H}_N
        \right],
 \end{equation}
 which has a form of {\em Ising type exchange interaction} for $N$ qubits.
 For other terms of the $N$-qubit interaction,
 the coefficients of the terms can be called
 {\em $N$-qubit tunnel-exchange coupling constants},
 e.g.,
 $$J^{(N)}_{x\cdots y \cdots z}
  = C_{x\cdots y\cdots z}
  = \frac{1}{2^N} \mathrm{Tr}\left[
       \mbox{\boldmath $\sigma$}^{x}_1 \otimes
       \cdots 
       \mbox{\boldmath $\sigma$}^{y}_i 
       \cdots \otimes
       \mbox{\boldmath $\sigma$}^{z}_N \; {\cal H}_N
        \right],$$
 since the off-diagonal components of the Hamiltonian matrix
 result from the hopping (tunneling)
 between the sites (states).

 {\it Two qubit systems.}$-$
 For two qubit systems,
 the two-qubit interaction is given by
 $ H^{(12)}_2= \sum_{j\in \{x, y,z\}}
      J^{(2)}_{jj} \mbox{\boldmath $\sigma$}^{j}_1 \otimes
       \mbox{\boldmath $\sigma$}^{j}_2,$
 where
  $J^{(2)}_{xx}=-(t^{a}_2+t^{b}_2)/2$, $J^{(2)}_{yy}=(t^{a}_2-t^{b}_2)/2$,
  and $J^{(2)}_{zz}=
   \left( E_{\uparrow\uparrow} -E_{\uparrow\downarrow}
   -E_{\downarrow\uparrow} +E_{\downarrow\downarrow} \right) /4$.
 The two-qubit tunneling amplitudes, (i) $t^a_2$ and (ii) $t^b_2$,
 describe the tunneling processes,
 (i) $\left|\uparrow\uparrow \right\rangle \Longleftrightarrow
      \left|\downarrow\downarrow \right\rangle $
 in the parallel pseudo-spin states
 and
 (ii) $\left|\uparrow\downarrow \right\rangle \Longleftrightarrow
      \left|\downarrow\uparrow \right\rangle $
 in the anti-parallel pseudo-spin states.
 As expected, the exchange coupling constant $J^{(2)}_{zz}$
 is the energy difference between the parallel
 and anti-parallel pseudo-spin states.
 The two-qubit tunnelings contribute to
 the pseudo-spin exchange interaction.
 Then, $H^{(12)}_2$ has a form of XYZ model for two-pseudo spins.
 $t^a_2\ll t^b_2$ gives an XXZ pseudo-spin model
 and,
 for $J^{(2)}_{zz}=0$, i.e.,
   $E_{\uparrow\uparrow} +E_{\downarrow\downarrow}
    = E_{\uparrow\downarrow} +E_{\downarrow\uparrow} $,
 an XY pseudo-spin model.
 For $t^{a,b}_2 \ll J^{(2)}_{zz}$,
 $H^{(12)}_2$ becomes an Ising pseudo-spin model.
 This shows that various types of pseudo-spin models can be realized
 by manipulating the system parameters.

 {\it Three qubit systems.}$-$
 Next, for comparison, let us consider a two-qubit interaction
 of three qubit system given by
 $H^{(12)}_2= \sum_{j\in \{x, y,z\}}
      J^{(2)}_{jj0} \mbox{\boldmath $\sigma$}^{j}_1
              \otimes \mbox{\boldmath $\sigma$}^{j}_2
      +J^{(2)}_{xz0} \mbox{\boldmath $\sigma$}^{x}_1
              \otimes \mbox{\boldmath $\sigma$}^{z}_2
      +J^{(2)}_{zx0} \mbox{\boldmath $\sigma$}^{z}_1
              \otimes \mbox{\boldmath $\sigma$}^{x}_2$,
 where
 $J^{(2)}_{xx0}
  =-(\bar t^a_2+\bar t^b_2)/4-(\underline t^a_2+\underline t^b_2)/4$,
 $J^{(2)}_{yy0}
 =(\bar t^a_2-\bar t^b_2)/4+(\underline t^a_2-\underline t^b_2)/4$,
 and
 $J^{(2)}_{zz0} =(
    E_{\uparrow\uparrow\uparrow}
   -E_{\uparrow\downarrow\uparrow}
   -E_{\downarrow\uparrow\uparrow}
   +E_{\downarrow\downarrow\uparrow} )/8
   +(
   E_{\uparrow\uparrow\downarrow}
   -E_{\uparrow\downarrow\downarrow}
   -E_{\downarrow\uparrow\downarrow}
   +E_{\downarrow\downarrow\downarrow}
  )/8$.
 Here, $\bar t$($\underline{t}$) denote the two-qubit tunnelings
 for the up(down) state of  the third pseudo-spin.
 Compared to the two-qubit interaction in two-qubit systems, interestingly,
 there are the two extra tunnel-exchange coupling terms,
 $J^{(2)}_{xz0}$ and $J^{(2)}_{zx0}$, mediated by the single-qubit
 tunnelings.

 In the three-qubit interaction $H^{(123)}_3$, the three-qubit
 exchange coupling constant is given by
  $J^{(3)}_{zzz} =(
    E_{\uparrow\uparrow\uparrow}
   -E_{\uparrow\downarrow\uparrow}
   -E_{\downarrow\uparrow\uparrow}
   +E_{\downarrow\downarrow\uparrow} )/8
   -(
   E_{\uparrow\uparrow\downarrow}
   -E_{\uparrow\downarrow\downarrow}
   -E_{\downarrow\uparrow\downarrow}
   +E_{\downarrow\downarrow\downarrow}
  )/8$.
 Also, the single- and two-qubit tunnelings as well as the three-qubit
 tunnelings give rise to the three-qubit tunnel-exchange coupling
 constants.
 Especially, if the three-qubit tunnelings
 are stronger than the two-qubit tunnelings,
 the ground state can be in a Greenberger-Horne-Zeilinger (GHZ)
 and if the two-qubit tunnelings
 are stronger than the three-qubit tunnelings,
 a W-state can be generated in an excited state \cite{Kim07}.

 \begin{figure}
 \vspace{3.1cm}
 \includegraphics{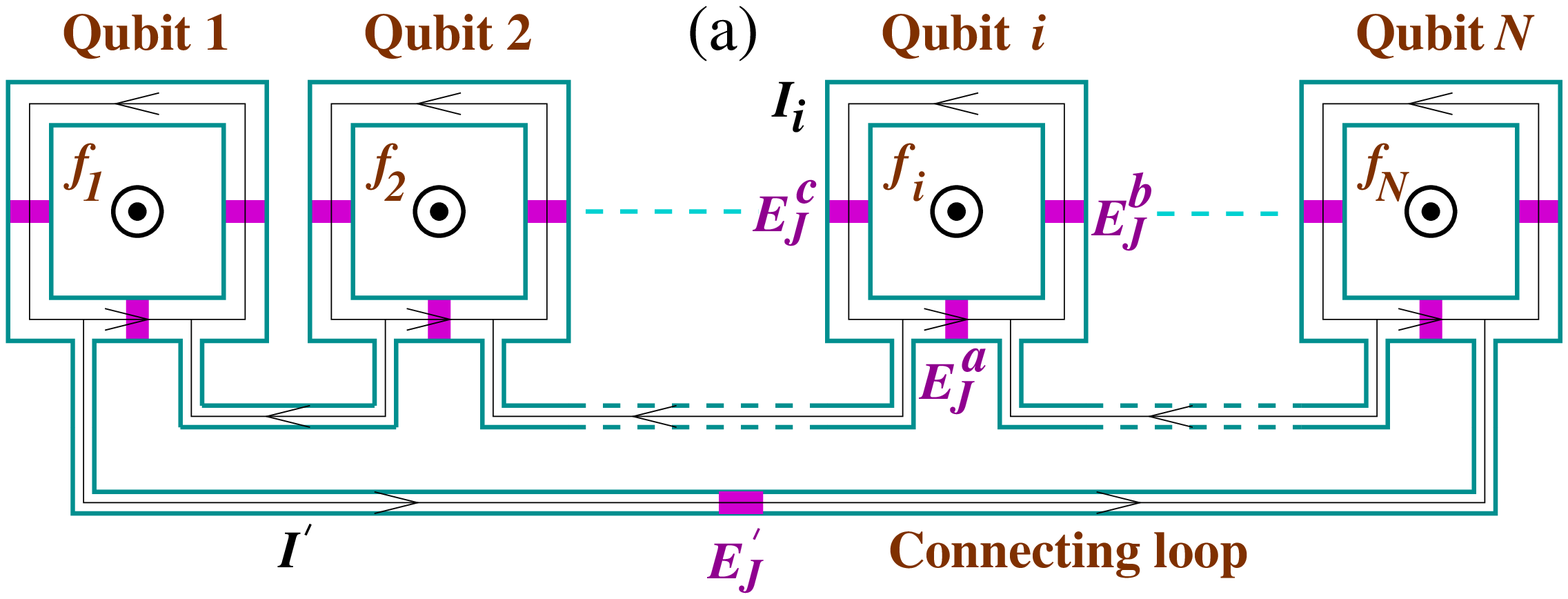}
 \vspace{4.0cm}
 \includegraphics{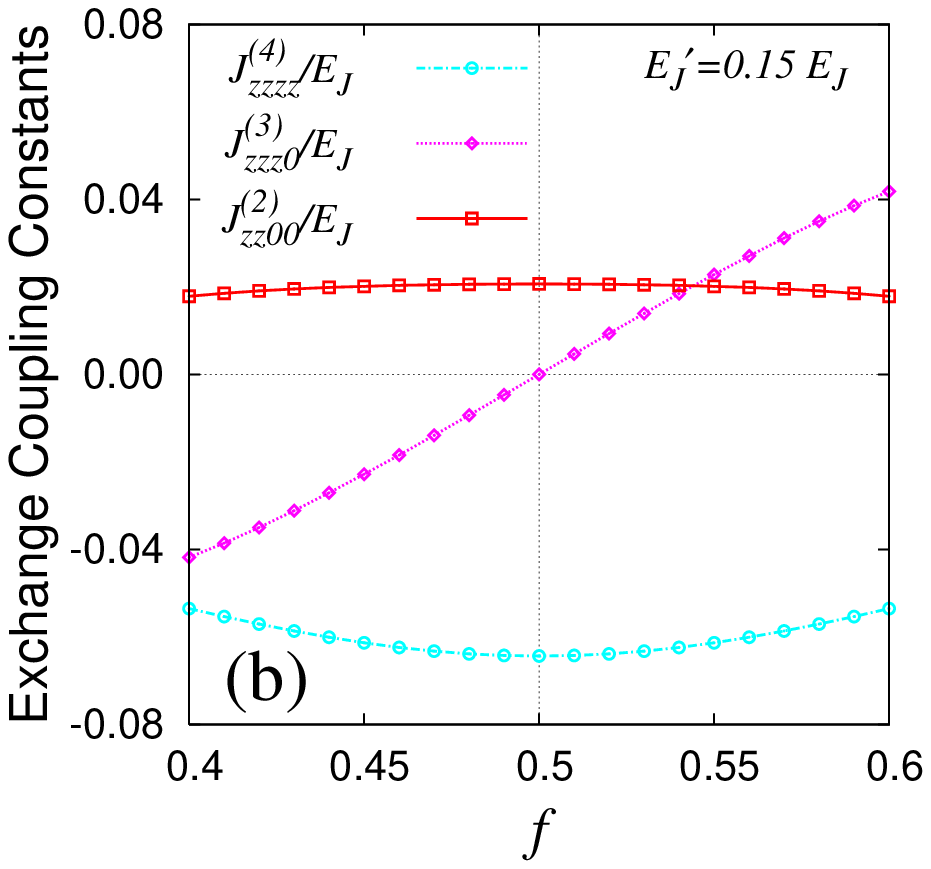}
 \includegraphics{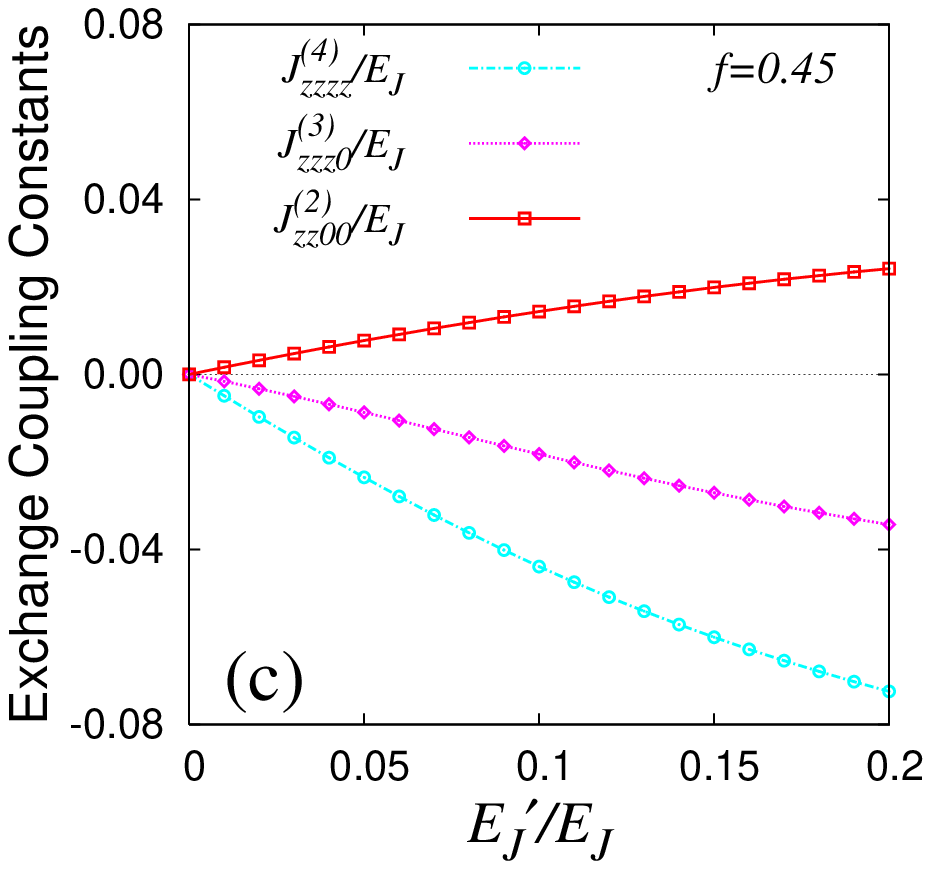}
 \vspace*{0cm}
 \caption{(Color online)
  Top: (a) A $N$ flux qubit system with one connecting loop.
  The $N$ superconducting loops are
  connected by a connecting loop interrupted by a Josephson junction,
  $E'_J$.
  In each qubit loop,
  the diamagnetic (paramagnetic) current states assigned
  by $\left|\downarrow\right\rangle (\left|\uparrow\right\rangle)$,
  are superposed, which makes the loop being regarded as a qubit.
  $\odot$ (oppositely $\otimes$) denote the directions of
  the applied and induced magnetic fields, $f_{i}=\Phi_{i}/\Phi_0$,
  in the qubit loop $i$.
  $I_i (I')$ stand for the currents
  in the qubit $i$ (connecting) loop.
  $E_{J}$'s are the Josephson coupling energies of the Josephson junctions
  in the connecting and qubit loops.
  The fluxoid quantization in the connecting loop
  gives rise to the boundary condition connecting
  the phases, $\varphi_i$, across each Josephson junction.
  Both the mutual inductances and the fluxoid quantization make
  it possible to realize many-qubit interactions in the $N$ flux qubit system.
  The many-qubit interactions are defined in the text.
  Bottom:
  Multiple qubit exchange coupling constants,
  $J^{(4)}_{zzzz}$, $J^{(3)}_{zzz0}$, and $J^{(2)}_{zz00}$,
  in the four qubits ($N=4$)
  as a function of
  (b) the applied magnetic field $f=f^{(i)}_{\rm ext}$ ($i=1,\cdots,4$)
      for $E'_J=0.15 E_J$ and
  (c) the Josephson energy $E'_J$ for $f=0.45$.
  Other parameters are
  $n_i=n'=0$ and $E^{\alpha}_J=E_J(\alpha\in\{a,b,c\})$.
  }
 \label{fig1}
 \end{figure}
 {\em Multiple qubit systems.}$-$
 To explore many-qubit interactions explicitly,
 let us consider a specific multiple-qubit system in Fig. \ref{fig1} (a).
 For simplicity, the inductances are assumed to be very small
 and then the inductive energy can be negligible.
 The boundary conditions for (i) the qubit loops and (ii) the connecting loop
 are reduced to
 (i) $ 2 \varphi^{}_i + \varphi^{a}_i = 2\pi (n_i-f^{(i)}_{\rm ext}) $
 and
 (ii) $ \varphi'= 2\pi n' - \sum^{N}_{i=1} \varphi^{a}_i$.
 The effective potential is given as 
 $U\!\!\! =\!\!\! \sum^{N}_{i=1}
      \left[ 4 E_J \sin^2 \frac{\varphi_i}{2}
      +2 E^a_J\sin^2\pi (n_i-f^{(i)}_{\rm ext}-\varphi_i/\pi) \right]
      +2 E'_J \sin^2 \pi \left[ n' -\sum^N_{i=1}(n_i-f^{(i)}_{\rm ext}
      -\varphi_i/\pi)\right].$

 For the four qubit system $(N=4)$,
 we plot the exchange coupling constants as a function of
 $f=f^{(i)}_{\rm ext}$ and $E'_J$
 in Fig. \ref{fig1} (b) and (c), respectively.
 At the co-resonance point, $f=0.5$, the three-qubit interaction
 disappears while the two- and four-qubit interaction strengths
 reach their maximum values in Fig. \ref{fig1} (b).
 The sign of the three-qubit interaction is changed from
 negative for $f < 0.5$ to positive for $f > 0.5$.
 As $E'_J$ increases, the two-, three-, and four-qubit interactions
 increase monotonically in Fig. \ref{fig1} (c).

 Interestingly, the four-qubit interaction is stronger than
 the two- and three-qubit interactions.
 That is, $J^{(4)}_{zzzz} \approx 3 J^{(2)}_{zz00} $.
 Also, the three-qubit interaction can be stronger than
 the two-qubit interaction for a certain applied magnetic field.
 This result seems to be counterintuitive.
 However, for an $N$ qubit system,
 the result can be understood from Eq. (\ref{fluxoid}) and
 the boundary condition of the connecting loop, without the assumption,
 $ L'_K I'/\Phi_0 =
  n'-(1/2\pi)( \varphi' + \sum^{N}_{i=1}
 \varphi^{a}_i)- f'_\mathrm{ind},
 $
 where $f'_\mathrm{ind}= L' I'/\Phi_0+\sum^N_{i=1} {\cal L}^{(i)}_M I_i/\Phi_0$
 with the mutual inductance ${\cal L}^{(i)}_M$.
 When one superconducting loop couples all qubit loops,
 all qubits are interconnected through the effective flux
 $f_{\rm eff} \equiv (1/2\pi)\sum^{N}_{i=1} \varphi^{a}_i$
 as well as  $f'_\mathrm{ind}$.
 Normally, the induced flux is much smaller than the effective flux, i.e.,
 $f^{(i)}_\mathrm{ind}, f'_\mathrm{ind} \ll f_{\rm eff}$, so that
 much stronger many-qubit interaction for the effective flux than
 for the induced magnetic flux can be expected.
 Therefore, if the $N$-qubit interaction is much stronger than
 other quit interactions,
 ${\cal H}_N \approx H^{(1\cdots N)}_N$ can map
 higher dimensional systems \cite{Yung}.

 We also considered two more models.
 (i) For $N$ qubits inductively coupled without
 any connecting loop, multiple qubit interactions
 are intrinsically involved but their strengths are very weak,
 for instance of $N=4$,
 $J^{(4)}_{zzzz} \approx 10^{-6} J^{(2)}_{zz00}$ in the parameters of Ref. \cite{Majer}.
 If the two-qubit interactions are much stronger than other multiple
 qubit interactions, 
 ${\cal H}_N \approx \sum_{i<j} H^{(ij)}_2$.
 Then, an $N$ qubits inductively coupled can be a many-body system
 in which one artificial-spin
 interact with all other artificial-spins by the two-body interactions.
 (ii) For the model of Ref. \cite{Kim},
 the multiple-qubit interactions
 behave similarly with the model of Fig. \ref{fig1} (a).
 In the same parameter values with Fig. \ref{fig1} (b), however,
 this model gives
 $J^{(4)}_{zzzz} \approx 0.17 J^{(2)}_{zz00}$.
 The two models show that the four-qubit interaction
 is smaller than the two-qubit interactions for the four qubit systems.
 In general, hence,
 many-qubit interactions are  dependent on
 specific experimental setups and varying the system parameters.
 Various types of artificial-spin systems can be prepared in flux
 qubit systems.
 Therefore, it is possible to explore a many-body system realized
 in flux qubit systems.

{\it Summary.}$-$
 We investigated many-qubit interactions in superconducting flux qubit systems.
 There are two types of many-qubit exchange interactions.
 One is similar with the Ising spin interaction,
 the other types of exchange interactions are due to
 macroscopic quantum tunnelings between the many-qubit states.
 Various types of many-qubit interactions can be realized experimentally in flux qubit systems.
 Moreover, an experimental setup can be provided to study
 many-body systems that can be mapped into many-flux qubit systems.

  Cho acknowledges the support from
  the Australian Research Council.

\end{document}